# RELATING ELECTRONIC STRUCTURE TO THERMOELECTRIC DEVICE PERFORMANCE

Shuaib Salamat, Abhijeet Paul, Mehdi Salmani-Jelodar, Changwook Jeong,

Gerhard Klimeck, and Mark Lundstrom

Network for Computational Nanotechnology

Birck Nanotechnology Center

Purdue University

West Lafayette, Indiana, 47907

Realistic thermoelectric modeling and simulation tools are needed to explain the experiments and for device design. In this paper, we present a simple computational technique to make use of rigorous band structure calculations in thermoelectric device design. Our, methodology, based on Landauer theory, provides a way to benchmark effective mass level models against rigorous band structures. It can also be used for determining the temperature-dependent Fermi-level from a given doping density and numerically generated density-of-states, which can then be used with the numerically generated distribution of channels to evaluate the temperature dependent electrical conductivity, thermopower, and electronic thermal conductivity. We illustrate the technique for silicon, for which well-calibrated band structure and transport data is available, but the technique is more generally suited to complex thermoelectric materials and provides an easy way to make use of first principles band structure calculations in device design.

## 1)    INTRODUCTION

The design of thermoelectric (TE) devices requires accurate device models that include electronic and heat transport as well as parasitics such as interface and contact resistances [1, 2]. The inputs to such models are thermoelectric transport parameters such as electrical conductivity, $\sigma(T)$, Seebeck coefficient, $S(T)$, and electronic and lattice thermal conductivity, $k_e(T), k_L(T)$. The precise determination of TE transport coefficients is, therefore, necessary for realistic device design and performance projections. The development of density functional theory [3, 4] opened up the doors for calculation of thermoelectric (TE) properties from first-principles. Although the techniques to evaluate TE parameters from first principles band structure calculations have evolved since then, they are still difficult to use and, therefore, rarely employed for device design and optimization. As a result, the effective mass approximation (EMA) is widely employed (often with first order corrections for band non-parabolicities). These handcrafted models can be difficult to develop and are rarely benchmarked against rigorous band structures. TE parameters can also be experimentally characterized, but the measured data is often too limited for comprehensive device design and optimization. Moreover, the assessment of new materials and structures for which measured data is limited should be done in a realistic device context.

Effective mass level models are generally satisfactory for parabolic bands but only for the energies close to the band edges. Concerns arise at high temperatures and high impurity concentration where the high energy, non-parabolic portion of the bands becomes important. Also, semiconductors with high $ZT$ have small band gaps, and non-parabolicity is an important factor in these materials [5-10]. Typically a Kane dispersion is used to handle non-parabolicity of the energy bands, which only works well in direct band gap semiconductors [11]. Much of the previous work largely relied on extending the EMA models to include the non-parabolicities to the first order [5-10, 12, 13] and typically treated scattering in power-law form [14-16], but constant relaxation time descriptions of scattering are also common [7, 8, 12, 17]. It is not clear, however, how well these first-order corrections correspond to the actual band structure, and the resulting expressions for the transport parameters can become quite complex with many parameters to keep track of. Due to large number of fitting parameters in the scattering models, experimental data is readily fit, but the physical soundness of the model is not clear. Therefore, it would be rather preferable to start with an accurate description of the band structure.

Accurate band structures are readily available from *ab initio* simulation codes [17-25], and many materials, including common thermoelectric materials, such as $Bi_2Te_3$, PbTe and PbSe, have been parameterized in nearest-neighbor tight binding models by different groups with different level of approximations [26-29]. Thermoelectric coefficients can be evaluated with first principle band structures, but very simple treatments of scattering (e.g. relaxation time approximation) are commonly-used [17-22, 24, 25, 30]. More advanced models, wherein scattering rates are computed from the first-principle band structure, are being developed [31, 32] but they are still not suitable for device design and analysis.

Recently, Jeong *et al.* presented a technique to compute TE parameters from full band dispersion using Landauer framework [33]. From the full band structure, a distribution of conducting channels, $M(E)$, is extracted. With descriptions of the energy-dependent mean-free-path for backscattering, the TE parameters are then readily evaluated. Extending Jeong's approach, this paper makes use of an advanced mode computing algorithm [34]. Besides $M(E)$, in order to evaluate the temperature dependent TE parameters, the temperature dependent Fermi-level, $E_F(T)$, is calculated from numerical density-of-states, $D(E)$, obtained from the full band $E(k)$. All calculation are done for Si, not because it is a good thermoelectric material but because well-calibrated band structure calculations are available [23, 28, 29, 35] and experimental data, especially for conductivity is abundant [35-38]. The specific results for Si described in this paper illustrate an excellent way to illustrate a more generally applicable technique. The specific contributions of this paper are the presentation of a clear way to quantitatively measure the errors associated with use of EMA models while at the same time outlining a generally applicable methodology to make use of full band structure descriptions for realistic TE material evaluation and device design.

The paper is organized as follows. In Sec. 2, a brief summary of the Landauer formalism, written in the form of differential conductance, is presented. In Sec. 3, tight binding simulation results are presented for the conduction and valence bands of silicon (Si). We first examine ballistic transport. Although unrealistic for traditional thermoelectric devices, the ballistic TE parameters depend only on band structure, so the fundamental differences between full band and EMA models can be isolated. Later on a realistic (but phenomenological) description of scattering is included to show that experimental data can be fit over a wide range of temperatures. Sec. 4 discusses the results within the Landauer framework, and also compares the full band approach to the EMA approach. Our conclusions are summarized in Sec. 5.

## 2) APPROACH

The standard expressions for thermoelectric parameters (electrical conductance ($G$), Seebeck coefficient ($S$) and electronic thermal conductance in zero voltage difference limit ($K_0$), all written in terms of the differential conductance, are

$$G = \int_{E_1}^{E_2} G'(E) dE \tag{1}$$

$$S = -\left(\frac{k_B}{q}\right) \frac{\int_{E_1}^{E_2} \left(\frac{E - E_F}{k_B T}\right) G'(E) dE}{G} \tag{2}$$

$$K_0 = T\left(\frac{k_B}{q}\right)^2 \frac{\int_{E_1}^{E_2} \left(\frac{E - E_F}{k_B T}\right)^2 G'(E) dE}{G} \tag{3}$$

$$K_n = K_0 - S^2 T G, \tag{4}$$

where $K_n$ is the electronic thermal conductance at zero electrical current. The limits of integration are from deep $(12 k_B T)$ in the valance band to deep in the conduction band, so the contribution of both bands are included in these expressions (bipolar effects). The differential conductance, $G'(E)$ relates the TE coefficients to the band structure as

$$G'(E) = \frac{2q^2}{h} M(E) \left(\frac{\ll \lambda(E) \gg}{\ll \lambda(E) \gg + L}\right) \left(-\frac{\partial f_0}{\partial E}\right), \tag{5}$$

where $M(E)$ is the number of conducting channels at energy, $E$. The term, $\ll \lambda(E) \gg / (\ll \lambda(E) \gg + L)$, is the transmission, $T(E)$, and $\ll \lambda(E) \gg$ is the average mean-free-path for backscattering and can be expressed phenomenologically in power law form as $\ll \lambda(E) \gg = \lambda_0 (E/k_B T)^r$, [33] where, $r$ is a characteristic scattering exponent specific to a scattering mechanism.

Traditionally, in Boltzmann transport approach, the differential conductance, $G'(E)$, is written in terms of the so-called transport distribution function (TDF), $\Sigma(E)$, which, in terms of Landauer parameters can be written as [33],

$$\Sigma(E) = \frac{2L}{h} M(E) T(E) \tag{6}$$

Although the two approaches are mathematically equivalent, the advantage of Landauer approach is that it separates the TDF into two parts, $M(E)$ that depends only on band structure and $T(E)$ that depends both on scattering and the band structure. The distribution of modes is formally related to the band structure as [33, 39],

$$M(E) = \frac{h}{2L} \sum_k |v_x| \delta(E - E_k), \tag{7}$$

where $|v_x| = \frac{1}{\hbar} \frac{dE}{dk}$ is the group velocity in the transport direction. For an effective mass model in 3D, $M(E)$ is given by [40]

$$M(E) = \frac{m^*_{DOM}}{2\pi\hbar^2}(E - E_c), \tag{8}$$

where $m^*_{DOM}$ is "distribution of modes effective mass". If the band is spherical, this mass is a single isotropic effective mass. For ellipsoidal energy surfaces with transport in the x-direction, it is given by $m^*_{DOM} = \sqrt{m^*_y m^*_z}$, for each equivalent ellipsoid. For the conduction band of Si, $m^*_{DOM} = 2m^*_t + 4\sqrt{m^*_l m^*_t} = 2.05 m_0$ and for the valance band of Si, $m^*_{DOM} = m^*_{hh} + m^*_{lh} = 0.59 m_0$ [33]. As will be seen in the Sec. 4, the analytical effective masses computed above do not necessarily provide a good fit to the full band $M(E)$ in the energy range of interest. Consequently, the fitted effective masses are defined.

The density of states in the effective mass model is given as [41],

$$D(E) = \Omega \frac{m^*_{DOS}(\sqrt{2m^*_{DOS}(E - E_c)}}{2\pi^2\hbar^3}, \tag{9}$$

where $m_{DOS}^*$ is the density of states effective mass. For the conduction band of Si, where the constant energy surfaces are ellipsoidal, $m_{DOS}^* = 6^{\frac{2}{3}}(m_l^* m_t^{*2})^{\left(\frac{1}{3}\right)} = 1.062m_0$, and for the valance band in Si with spherical constant-energy surfaces, $m_{DOS}^* = \left(m_{hh}^{\frac{3}{2}} + m_{lh}^{\frac{3}{2}}\right)^{\frac{2}{3}} = 0.55m_0$ [42]. Here too, it is found that 'fitted density-of-states effective masses' improve the accuracy, but the best procedure is to simply use the full band numerically generated $M(E)$ and $D(E)$.

For evaluating $M(E)$ and $D(E)$, beyond EMA, an accurate description of the electronic structure is needed. For the purpose of this paper, a well-calibrated sp$^3$d$^5$s$^*$ tight binding model is used [28, 29]. Extending Jeong's simple technique for extracting $M(E)$ from numerical table of $E(k)$ [33] we have developed a robust algorithm to calculate $D(E)$ besides $M(E)$, both with high computation speed and accuracy. The algorithm can treat 1D, 2D, or 3D structures; details can be found in Ref. [34]. After computing $M(E)$ and $D(E)$ from the given dispersion, three thermoelectric parameters [Eqs. (1)-(4)] can be evaluated, if $\ll \lambda(E) \gg$ is known. For ballistic transport, $T(E) = 1$, and the conductance becomes independent of length and only depends on the band structure. As will be shown later, the ballistic TE parameters provide a test of the accuracy of effective mass model against a full band structure. To describe real devices, however, scattering needs to be included. As will be discussed in Sec. 4, a physics-based but phenomenological approach of electron scattering that describes the mean-free-paths for each relevant scattering mechanism in a power law form is used.

The approach described above allows the computation of TE parameters for a given temperature ($T$) and Fermi-level ($E_F$). In practice, however, a material sample is doped to a certain density and hence both the Fermi-level and TE coefficients much be calculated as a function of temperature. The temperature dependent Fermi-level, $E_F(T)$, is determined by the charge neutrality relationship [42],

$$p_0 - n_0 + N_D^+ - N_A^- = 0, \tag{10}$$

where $p_0$ and $n_0$ are hole and electron concentrations, $N_D^+$ and $N_A^-$ are ionized donor and acceptor concentrations. For the full band calculations Eq. (10) modifies to,

$$\int_{E_1}^{E_v} D(E)\big(1 - f(E,T)\big)dE - \int_{E_c}^{E_2} D(E)f(E,T)dE + N_D^+(T) - N_A^-(T) = 0, \tag{11}$$

where $f(E,T)$ is Fermi function. In the EMA, the charge neutrality equation [Eq. (10)] simplifies to,

$$N_V(T)\mathcal{F}_{\frac{1}{2}}(\eta_V) - N_C(T)\mathcal{F}_{\frac{1}{2}}(\eta_C) + N_D^+(T) - N_A^-(T) = 0, \tag{12}$$

where $N_C$ and $N_V$ are effective density of states in conduction and valance bands, $\mathcal{F}_{\frac{1}{2}}$ is Fermi-dirac integral of order one-half and, $\eta_V$ and $\eta_C$ are given by, $\eta_V = (E_V - E_F)/k_B T$ and $\eta_C = (E_F - E_C)/k_B T$.

Using $D(E)$ from the full band structure, the charge neutrality relationship in Eq (11) is solved for $E_F(T)$ using "Trust-region" algorithm, which is a well-established technique in non-linear optimization problems for assuring fast convergence [43, 44]. The band gap variation with temperature and partial ionization of dopants are included in the calculations. The limits of the integrals $E_1$ and $E_2$ [in Eq. (1)-(3) and Eq. (11)] are chosen carefully so as to include all the significant energy contributions (99.99% accuracy) to the integrands. The electronic heat conduction has the largest contribution deep in the band. For $E_F$ at the band edge, placing $E_1$ and $E_2$ $12 k_B T$ into the conduction band (or valance band) is sufficient for all integrals [34].

In brief the approach is as follows. First, the electron dispersion relation $E(k)$ is computed. For this work, the sp$^3$d$^5$s$^*$ tight binding model with spin-orbit coupling is used [28, 29]. This model is utilized only to illustrate the technique which can be employed equally well with other band structure descriptions. Second, $M(E)$ and $D(E)$ are extracted from $E(k)$. Third, a comparison of full band and EMA based ballistic TE parameters vs. $E_F$ at 300 K and at high temperature, 900 K is presented. This comparison establishes the amount of the error, between the full band and EMA calculations that is associated with $M(E)$. Next Eq. (11) is solved iteratively for $E_F$ as a function of temperature and impurity concentration (for the EMA Eq. (12) is solved similarly). Once $E_F(T)$ is determined, the ballistic TE parameters vs. temperature at different impurity concentrations are calculated and comparison of EMA to full band is made to establish the additional part of the error due to $D(E)$. Finally we show that measured data can be fit with a physics based scattering model and full band $M(E)$ and $D(E)$.

3)      **RESULTS**

In this section, the techniques described in Sec. 2 are illustrated and a comparison of full band and EMA results over a range of doping densities and temperatures is made. Ballistic transport is assumed, so the results provide a clear comparison of the band structure models. In the process, it is also demonstrated that full band calculations are easily done and well suited to device design and optimization. The material of choice is Si; its conduction band is approximately parabolic, but the valance band is highly warped [42].

Figure 1 shows the tight binding band structure for Si used in this work [28, 29]. High symmetry points are indicated on the figure. The conduction and valance bands are composed of a number of bands and the bands exhibit local and absolute minima at the zone center or along the high symmetry directions. The effective masses are calculated around the global band minima and band maxima for conduction band and valance band, respectively (conduction band minima is along $< 100 >$ direction in the $k$-space and valance band maxima is at the zone center i.e.$\Gamma$-point). As expected, the values of density of states effective masses computed from our tight binding model agree closely with the values reported in literature [42, 45] (compare tight binding values, for conduction band: $m_l^* = 0.8539m_0$, $m_t^* = 0.1962m_0$ and $m_{DOS}^*|_{CB} = 1.062m_0$, for valance band: $m_{hh} = 0.49m_0$ $m_{lh} = 0.17m_0$ and $m_{DOS}^*|_{VB} = 0.55m_0$ with Ref. [42] page 81 and page 96) .

Figure 2 shows three calculations of $M(E)$ for Si conduction and valance band. The first is a numerical calculation obtained directly from the $E(k)$ in Fig. 1Figure 1 using procedure given in Ref. [34]. The second is the analytical calculation using the effective masses extracted from the $E(k)$ at the band minima (maxima). Finally, we show a calculation using the distribution of modes effective mass as a fitting parameter. The fitting range was chosen to be $12k_BT$, where $T$ is the maximum working temperature (this is the energy range wherein all integrands in the thermoelectric coefficients converge to a 99.99% accuracy). As seen in Fig. 2, $M(E)$ for both the conduction band and valance band increase linearly with energy. A small nonlinearity in the conduction band $M(E)$ can be seen. This is a contribution from a second conduction band at X (zone boundary along $< 100 >$ direction in $k$-space) and 0.12 eV above the band minima. The analytical expression (dashed line) provides a good fit only very near the band edge. Adjusting the effective mass produces a reasonable fit to the conduction band $M(E)$ over the energy range of interest. Surprisingly, the highly warped valence band also displays the linearly increasing $M(E)$ expected for a parabolic band, but the analytically computed value is much different. Still, if one uses the modes effective mass as a fitting parameter, good results are obtained.

Figure 3 shows three calculations of $D(E)$ for Si conduction and valance band. Again, the first is the full band calculation, the second is the analytical calculation using effective masses extracted from the full band $E(k)$, and the third uses the density-of-states effective masses as fitting parameters. The fitting range is again $12k_BT$, as it was for $M(E)$. Examining Figs. 2 and 3 closely, it is interesting to note that conduction band non-parabolicity is much more apparent in $D(E)$ than in $M(E)$. This occurs because $M(E)$ is proportional to velocity times density-of-states [41]; non-parabolicity increases $D(E)$, but it decreases the velocity, so the effects partially offset. Again, it is surprising to see that the EMA fit is a better approximation for the valance band $D(E)$ than for the conduction band $D(E)$, despite the warped nature of the valance band.

Differences in the TE parameters from the full band and EMA band structures are due to the differences in both $M(E)$ and $D(E)$. It is rather straight forward to isolate the error associated with each. If we compute TE coefficients at the same $E_F$ for a given temperature, the error is solely due to $M(E)$. Figure 4, shows the computed TE parameters w.r.t. $E_F$ at $300K$. The results for 300 K and 900 K are also reproduced in Table 1. As seen in Fig. 4 and Table 1, at the same $E_F$, EMA provides a reasonably good approximation for full band, with maximum associated error of 17% (Ref. Table 1, compare $m^*(fit)$ corresponding to $K_e$ for p-type doping).

Thermal and electrical properties of Si have been studied extensively [23, 35, 37, 38, 46], but few works in literature give thermoelectric properties as a function of doping and temperature and none systematically compare full band and EMA models. In this work, the $E_F(T)$ is calculated iteratively based on the $D(E)$ for different impurity concentrations and temperatures. This gives us the ability to study a wide range of doping over a range of temperatures (freeze out, extrinsic and intrinsic regime). Band gap variations with temperature, partial ionization of dopants at low temperature (carrier freeze out), and intrinsic carrier concentration at high temperatures have been accounted for in the calculations. Figure 5, shows the $E_F$ positioning in Si as a function of temperature and impurity concentration. The doping range is chosen from very low doping to degenerate doping. As the temperature increases $E_F$ of the doped sample converges to intrinsic $E_F$ as the carriers excited across the band gap exceed the fixed number of carriers derived from impurity concentration and the material becomes intrinsic (more prominent for low impurity levels in the studied temperature range, 100-900 K).

Figures 6 and 7 display the thermoelectric properties for n-type and p-type materials, for an impurity concentration of $10^{19}$ per cm$^3$ and temperature range of 100-900 K. Full band, analytical and fitted-effective-mass results are shown in these figures. The calculations were performed for a range of doping, but for clarity only one impurity concentration is shown. While the functional variation of most parameters is intuitively reasonable, the initial rise of Seebeck coefficient with temperature and subsequent drop at high temperatures necessities some explaining. The functional behavior of $S(T)$ can be understood within the context of Maxwell Boltzmann statistics, if we consider $S(T)$ as average energy of transport above $E_F$. For n-type material we have

$$S_n(T) = -\frac{k_B}{q}\left(\delta - \frac{(E_F - E_C)}{k_B T}\right),\tag{13}$$

where $\delta$ is the average energy above the conduction band edge at which the heat flows. It is important to note that both $E_F$ and $E_C$ are a function of temperature. As the temperature increases, the interplay between two parameters $(E_F - E_C)$ and $k_B T$ determines the outcome of $S(T)$. In the beginning $(E_F - E_C)$ increases more rapidly and, therefore, $S(T)$ increase. As the temperature continues to increase, rise in $(E_F - E_C)$ nearly saturates and the $k_B T$ term takes over, therefore, $S$ starts decreasing.

As was shown in Fig. 2, $M(E)$ is approximately linear with energy as expected for parabolic bands [33], so even when the bands are highly non-parabolic, it is relatively easy to fit $M(E)$ with effective mass like expressions. As discussed earlier, the differences between the full band and fitted EMA results for the TE parameters vs. $E_F$ is less than 17% (Ref. Table 1). Figure 3 shows, however, that the density of states is harder to fit with parabolic band expressions. As shown in Table 1, the maximum error for the temperature-dependent TE parameters can be large (102% associated with conductance ($G$) and 71% associated with electronic thermal conductivity ($K_e$) and power factor ($PF$). Better analytical expressions for $D(E)$ could be developed (e.g. non-parabolic energy bands, multiple bands etc.) but it is easier to just directly use the tables of $M(E)$ and $D(E)$ obtained from the full band structure calculations.

## 4) DISCUSSION

While $M(E)$ and $D(E)$ can be computed from the band structure, scattering is more complicated, but it is essential to include scattering in the analysis of TE devices. Although a first-principles treatment of scattering using Fermi's golden rule is possible, this is not practical for device design because of the computational burden. Though strictly valid for parabolic bands, power-law scattering expressions are commonly used and should be an acceptable empirical expedient so as to keep the methodology practically useful and tractable. The fact that $M(E)$ varies approximately linearly with energy, as expected for parabolic bands, lends some credence to the procedure.

Figure 8 shows the full band as well as EMA (fit) calculations along with experimental data. We consider acoustic phonon (AP) and ionized impurity (II) scattering in power-law form and obtain the overall mean free path using

$$\frac{1}{\lambda(E)} = \frac{1}{\lambda_{AP}(E)} + \frac{1}{\lambda_{II}(E)} \tag{14}$$

where, $\lambda_{AP} = \lambda_0^{AP}(E/k_BT)^r$, and $\lambda_{II} = \lambda_0^{II}(E/k_BT)^r$ and $r$ is the characteristic scattering exponent ($r = 0$ for AP scattering and $r = 0 - 2$ for II scattering)[47].

Since the scattering exponent for ionized impurity scattering varies between 0 and 2, we determine the scattering exponent by using the experimental conductivity vs. temperature plot and taking the slope in the low temperature regime. The scattering exponent thus obtained is $r = 5/4$. Incorporating the above scattering mechanisms, full band and EMA calculations are performed for the TE parameters. As seen in Fig. 8, full band calculation provides a good match to the experimental data.

It is important to note that experimental data could also possibly be matched by using effective scattering exponent in power-law form (which will account for both AP and II scattering mechanisms)

$$\lambda = \lambda_0 \left(\frac{E}{k_BT}\right)^r \tag{15}$$

This approach is not unreasonable and has been proposed as a satisfactory approximation for relaxation time processes by Goldsmid [48]. However, the scattering methodology utilized in this work is more plausible and provides physically meaningful mean-free-paths.

5) **SUMMARY AND CONCLUSION**

In this paper, we used the Landauer approach to calculate thermoelectric transport coefficients from a full band electronic dispersion. This approach relies on the calculation of the distribution of modes, $M(E)$, from a given band structure, and on the transmission, $T(E)$, which depends on band structure and scattering. The approach provides a way to test the accuracy of effective mass level models against full band dispersions. Errors as large as 100% were found even for a material like silicon with reasonably parabolic energy bands. For both conduction and valence bands, we find, however, that $M(E)$ generally increases approximately linearly as the energy with respect to the band edge increases. The band edge effective masses do not, provide a good fit to the numerically evaluated $M(E)$, but if the effective mass is used as a fitting parameter, reasonable results can be obtained. More complicated models including non-parabolicity and additional bands can provide better fits, but the best (and easiest) approach is to simply use the numerically generated tables for $M(E)$ and $D(E)$. Finally, we showed that power law descriptions of scattering process can be used with the full band $M(E)$ to accurately describe the temperature-dependent TE parameters. Equally good fits could be accomplished with simpler band structure models, but using an accurate band structure results in fitted mean-free-paths that have physical significance. The methodology presented in this paper not only provides direct test for the accuracy of simple band structure models, it provides a simple way to make use of first principles band structure calculations in thermoelectric device design and analysis.

**Acknowledgement**

S. Salamat would like to acknowledge Network for Computational Nanotechnology (NCN) for funding and computational support. A. Paul acknowledges support from FCRP under MSD and MIND. C. Jeong acknowledges support from SRC Focus Center in Material, Structure and Devices. We thank M. Fischetti for insight and discussion (with M. Lundstrom) on silicon band structure.

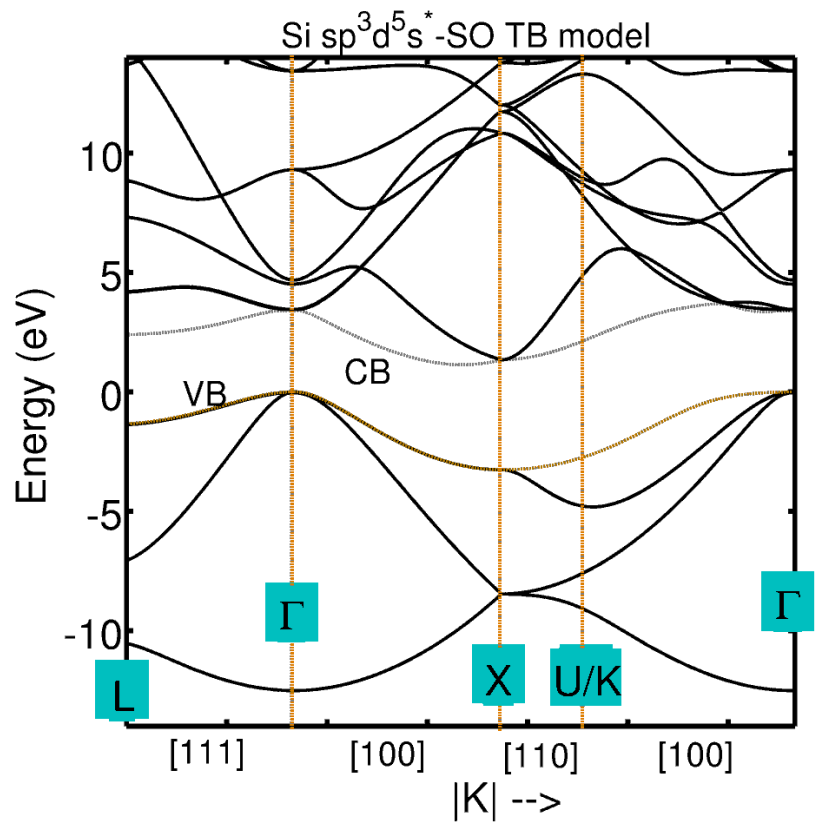

Figure 1. Energy-band structure ($E(K)$ diagram) characterizing the conduction and valance bands of Si [28, 49].

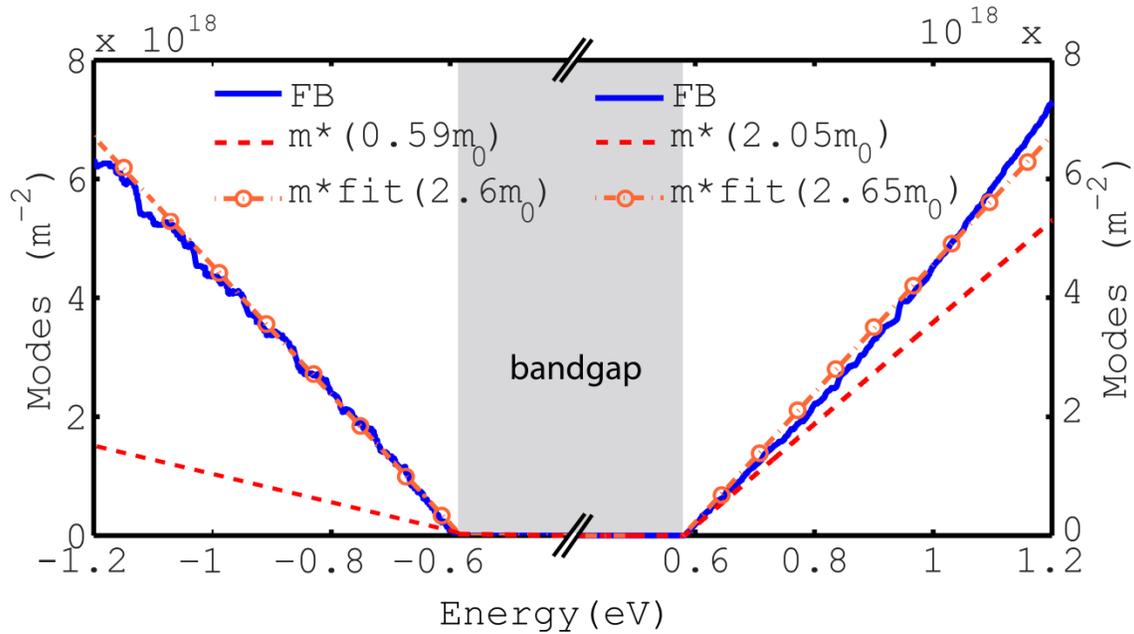

Figure 2. Plot showing three calculations of Modes for conduction and valance band of Si. First (solid line) is a numerical calculation obtained directly from the full band $E(k)$ in Figure 1. The second (dash line) is the analytical calculation using the effective masses extracted from the $E(k)$ at the band minima (maxima), third (dash-dot line with circles) uses the distribution of modes effective mass as a fitting parameter $m^*(fit)$ is extracted by fitting over the energy range of $12k_BT$. This is the energy range wherein the integral for TE coefficient converge to 99.99% accuracy.

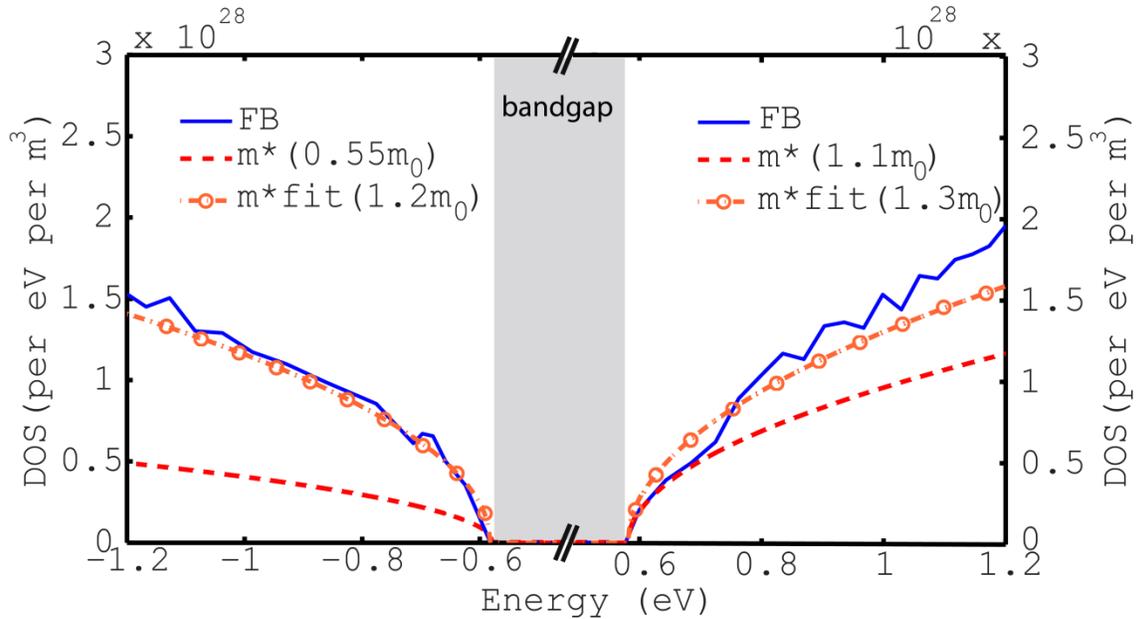

Figure 3. Plot showing three calculations of $D(E)$ for the conduction and the valance band of Si. First (solid line) is a numerical calculation obtained directly from the full band $E(k)$ in Figure 1. The second (dash line) is the analytical calculation using the effective masses extracted from the $E(k)$ at the band minima (maxima), third (dash-dot line with circles) uses the distribution of modes effective mass as a fitting parameter $m^*(fit)$ is extracted by fitting over the energy range of $12k_BT$. This is the energy range wherein the integral for TE coefficient converge to 99.99% accuracy.

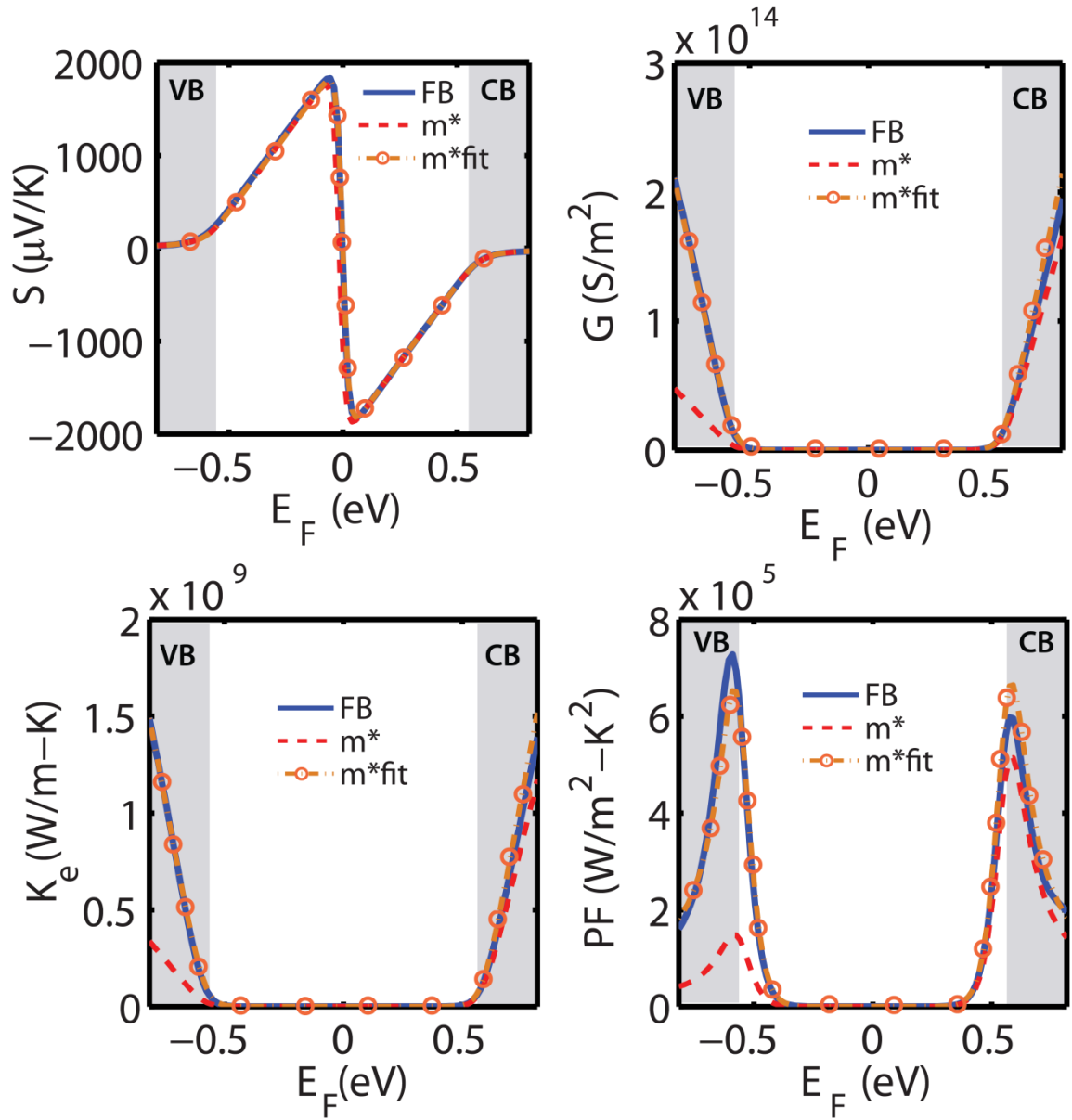

Figure 4. Thermoelectric transport coefficients as a function of Fermi-level. All calculation pertain to the temperature $300K$. Numerical values for 300K as well as 900K are reproduced in Table 1.

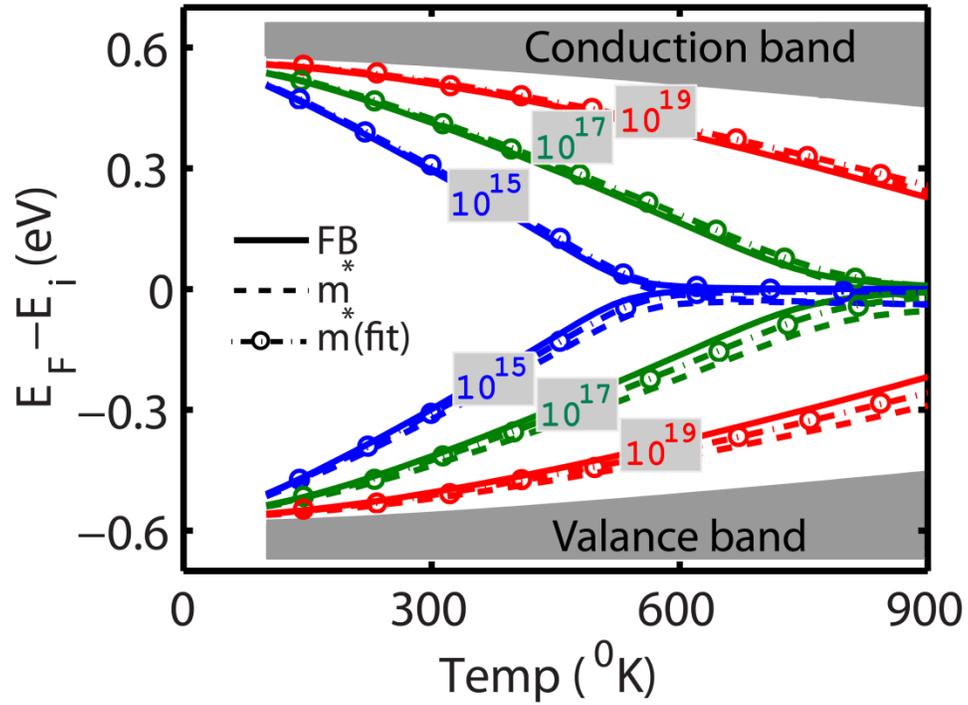

Figure 5. The Fermi-level positioning in Si as a function of temperature and impurity concentration. The dependence of band gap on the temperature is also shown. Partial ionization of dopants at low temperatures is taken into account in these calculations.

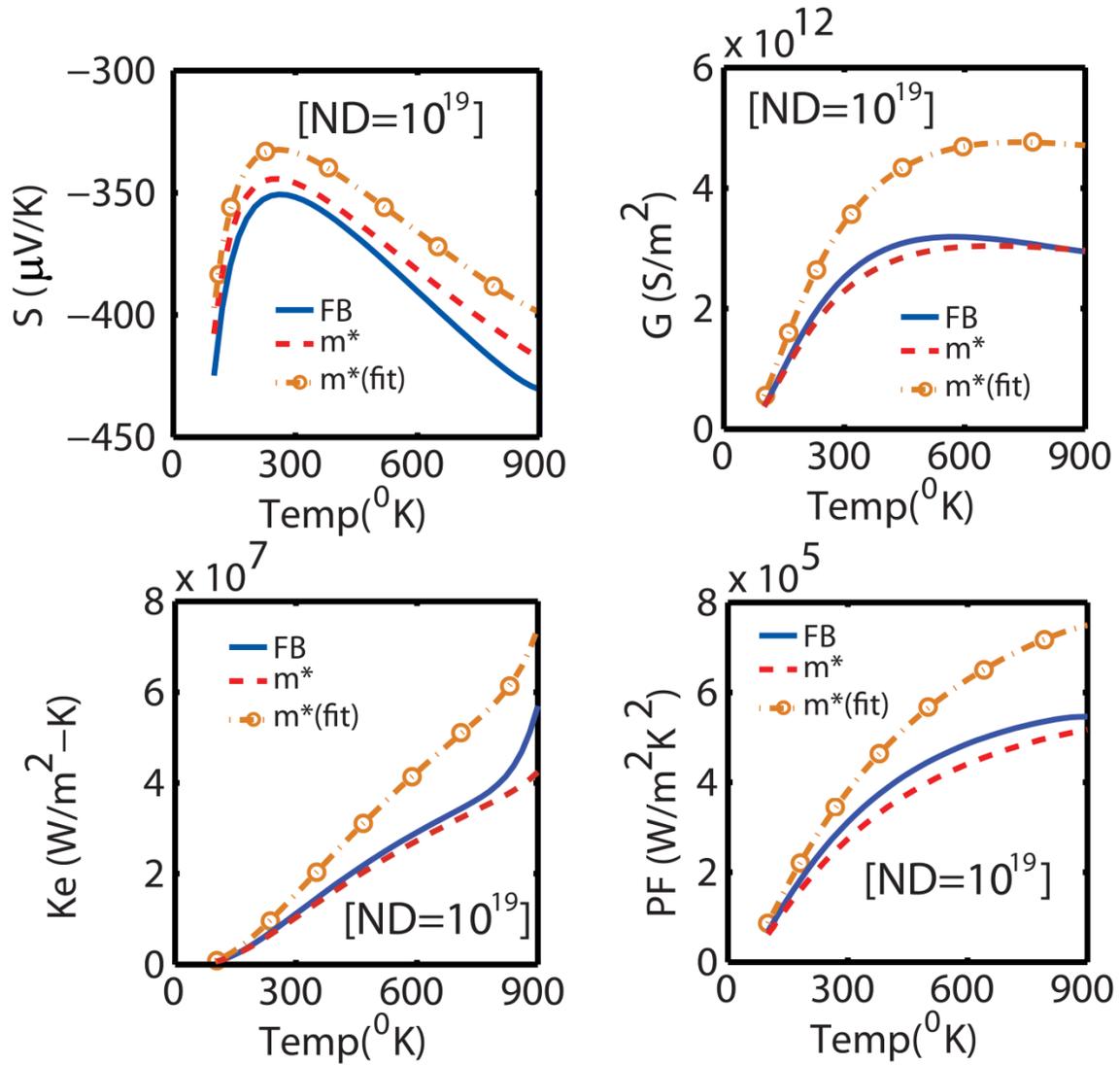

Figure 6. Thermoelectric coefficient for n-doped Si as a function of temperature. For clarity only one impurity concentration is shown. (a) Seebeck coefficient, (b) electrical conductance, (c) electronic thermal conductance, (d) power factor

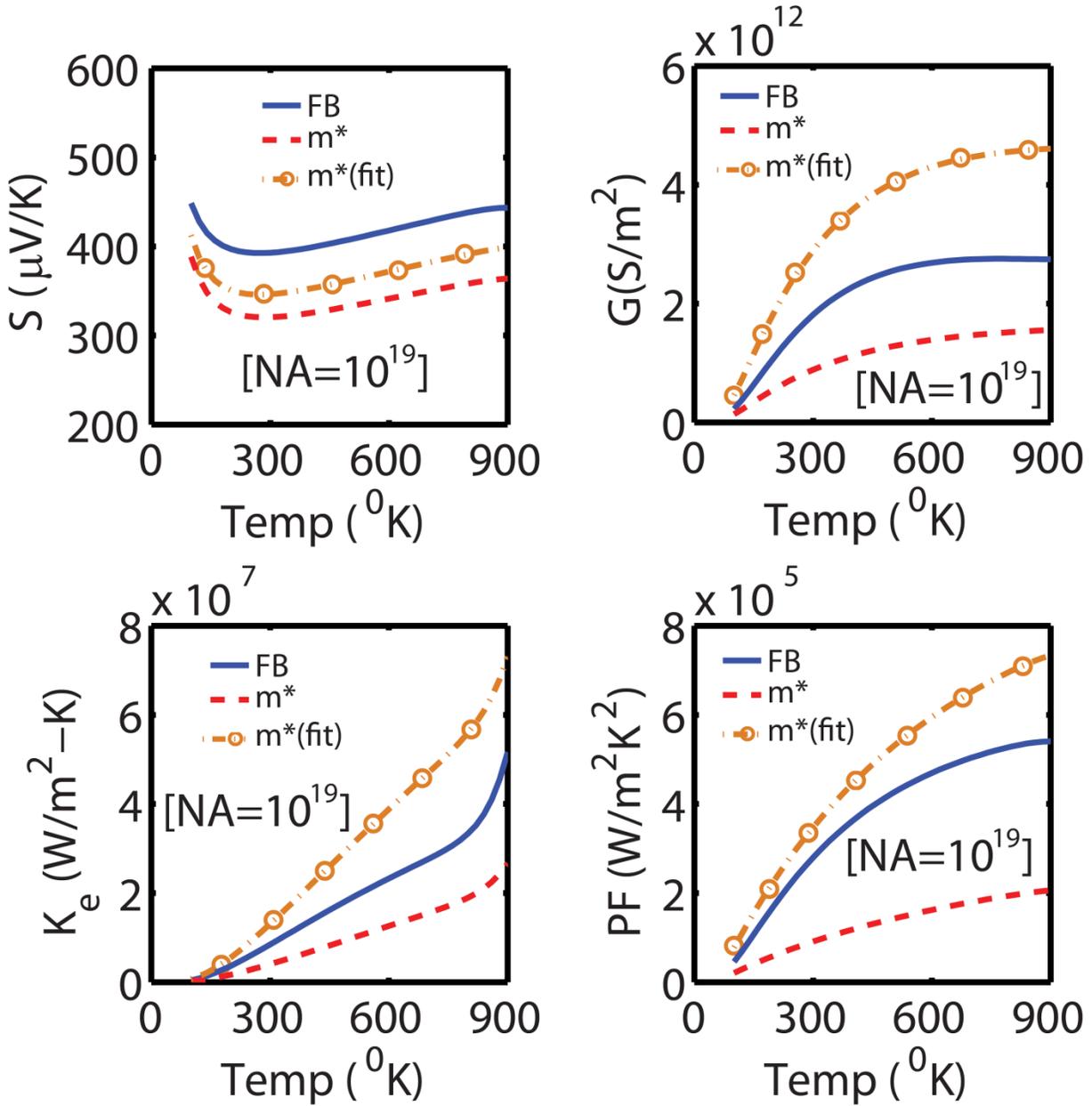

Figure 7. Thermoelectric coefficient for p-doped Si as a function of temperature. For clarity only one impurity concentration is shown. (a) Seebeck coefficient, (b) electrical conductance, (c) electronic thermal conductance, (d) power factor

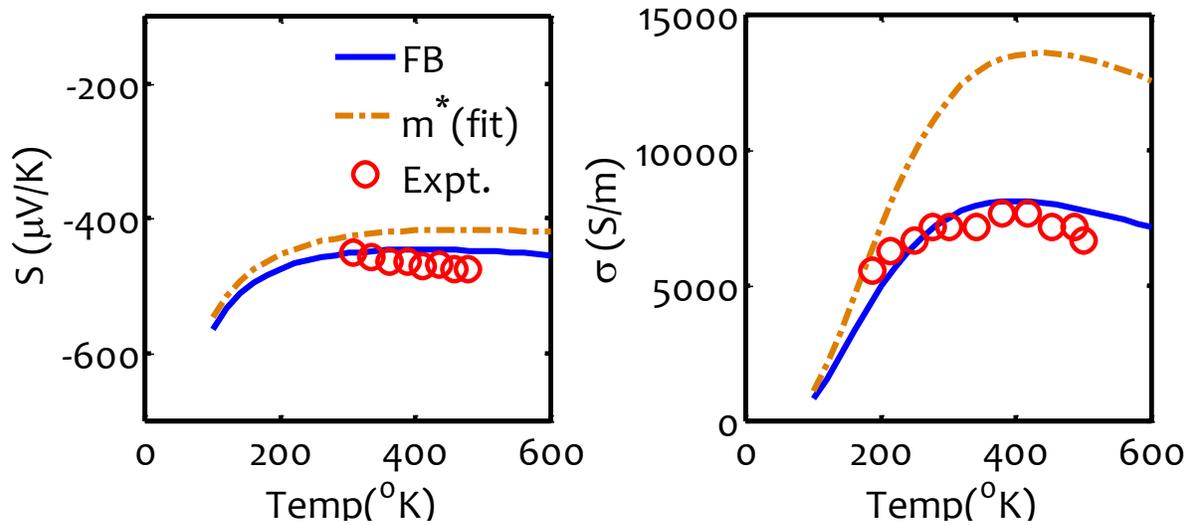

Figure 8. Seebeck coefficient and electrical conductivity for n-doped Si. Circles: experimental data from Ref. [46] for Seebeck coefficient corresponding to doping density of $10^{19}\ cm^{-3}$ and from Ref. [37] for electrical conductivity corresponding to doping density of $2.5 \times 10^{18} cm^{-3}$. Two scattering mechnisim, AP and II are considered with each assumed to be in Power-law form. Total $\lambda$ is calculated using Matheissen's rule ($\lambda_0^{AP} = 100nm, \lambda_0^{II} = 1nm, r_{AP} = 0, r_{II} = \frac{5}{4}$). The scattering exponent of II scattering is found by fitting the experimental data in low temperature regime.

Table 1. Errors in TE parameters of silicon. Errors in EMA (analytical) and EMA (fit) are calculated as a percentage of Full band results. Results are computed for the constant Fermi-level as well as for impurity concentration of $10^{19} \, cm^{-3}$ in the temperature range of $100K - 900K$. For the constant Fermi-level case, error is calculated at the $E_F$ that maximized PF.

| Parameter | Doping | Max error at same Ef at 300K (%age) | | Max error at same Ef at 900K (%age) | | Max total error at 300K (%age) | | Max total error 100K-900K (%age) | |
|---|---|---|---|---|---|---|---|---|---|
| | | $m^*$ | $m^*(fit)$ | $m^*$ | $m^*(fit)$ | $m^*$ | $m^*(fit)$ | $m^*$ | $m^*(fit)$ |
| $S$ | n-type | 1 | 1 | 2 | 2 | 2 | 5 | 4 | 7 |
| $G$ | -"- | 16 | 9 | 15 | 10 | 9 | 36 | 9 | 60 |
| $K_e$ | -"- | 14 | 10 | 20 | 4 | 7 | 39 | 26 | 48 |
| $PF$ | -"- | 14 | 12 | 19 | 6 | 12 | 22 | 13 | 37 |
| $S$ | p-type | 10 | 10 | 1 | 1 | 18 | 12 | 18 | 12 |
| $G$ | -"- | 74 | 16 | 77 | 1 | 51 | 60 | 51 | 102 |
| $K_e$ | -"- | 74 | 17 | 76 | 5 | 52 | 55 | 54 | 71 |
| $PF$ | -"- | 79 | 6 | 78 | 1 | 67 | 24 | 68 | 71 |